\documentclass[twocolumn,preprintnumbers,amsmath,amssymb]{revtex4}

\usepackage{graphicx}  
\begin{document}

\title{
Graphene vertical cascade interband terahertz and infrared photodetectors \textit{}}
\author{V. Ryzhii,$^{1,2}$ T. Otsuji,$^1$ M. Ryzhii,$^3$
V. Ya. Aleshkin,$^4$ A.~A.~Dubinov,$^{4}$\\ D. Svintsov,$^5$
 V. Mitin,$^6$
 and M. S. Shur$^7$ }
\address{$^1$Research Institute for Electrical Communication, Tohoku University,  Sendai 980-8577, Japan\\ 
$^2$ Center for Photonics and Infrared Engineering, Bauman Moscow State Technical University and 
Institute of Ultra High Frequency Semiconductor Electronics of RAS,
Moscow 105005, Russia\\
$^3$ Department of Computer Science and Engineering, University of Aizu, Aizu-Wakamatsu 965-8580, Japan\\
$^4$ Institute for Physics of Microstructures of RAS and Lobachevsky State University of Nizhny Novgorod, Nizhny Novgorod 603950, Russia\\
$^5$ Institute of Physics and Technology of RAS and  Department of General Physics, Moscow Institute of Physics and Technology, Dolgoprudny 141700, Russia\\
$^6$ Department of Electrical Engineering, University at Buffalo, SUNY, Buffalo, New York 1460-1920, USA\\
$^7$ Department of Electrical, Electronics, and System Engineering and Department of Physics and Astronomy, Rensselaer Polytechnic Institute, Troy, New York 12180, USA}

\begin{abstract} We propose and evaluate the   vertical cascade terahertz  and infrared  photodetectors based on  
  multiple-graphene-layer (GL) structures with thin  tunnel  barrier layers (made of tungsten disulfide or related materials).
The photodetector operation  is associated with   the cascaded radiative electron transitions
from the valence band in GLs to the conduction band  in the neighboring GLs (interband- and inter-GL transitions).
We calculate the spectral dependences of the responsivity and detectivity
for the vertical cascade interband GL- photodetectors (I-GLPDs) with different number of GLs and  doping levels at different bias voltages  in a wide temperature range.
  We show the possibility of an effective manipulation
of the spectral characteristics by the applied voltage. The spectral characteristics
depend also on the GL doping level that opens up the prospects of using I-GLPDs  in the multi-color systems. The advantages of I-GLPDs  under consideration are associated with their sensitivity to the normal incident radiation,
weak temperature dependence of the dark current as well as high speed of operation. The comparison of the proposed I-GLDs with the quantum-well intersubband photodectors 
demonstrates the superiority of the former, including a better detectivity at room
temperature and a higher speed. 
The vertical cascade I-GLDs can also surpass the lateral p-i-n GLDs in speed.
\end{abstract}

\maketitle
\newpage

\newpage
%%%%%%%%%%%%%%%%%%%%%%% References %%%%%%%%%%%%%%%%%%%%%%%%%

%%%%%%%%%%%%%%%%%%%%%%%%%%  body  %%%%%%%%%%%%%%%%%%%%%%%%%%
%\newpage
\section{Introduction}
The GL-based  heterostructures with the thin barrier layers made of Boron Nitride (hBN),
Tungsten Disulfide (WS$_2$), and other transition metal dichalcogenides have recently attracted a considerable interest. 
Several novel devices have been proposed and realized~[1-14].
Due to the gapless energy spectrum of the GLs in such structures (similar to that single-GLs and non-Bernal 
stacked twisted GLs in the multiple-GL structures~\cite{15}), such heterostructures can be used
in terahertz (THz) and infrared (IR) photodetectors. 
The GL-structures with the tunneling transparent inter-GL barriers can be particularly useful in the novel   
THz and IR photodetectors surpassing or complementing other GL-based 
photodetectors~[16-24] and photodetectors
based on more standard semiconductor materials~\cite{25}.

In this paper,
we propose and evaluate the vertical intersubband THz and IR photodetectors based on  the Bernal-stacked multiple-GL structures
with tunneling barrier layers using  the cascade of the interband inter-GL
radiative transitions  
The cascade of the electron tunneling and radiative-assisted tunneling
processes supports the vertical dark current and photocurrent between  the top and bottom GLs (playing the role of the emitter and collector). The advantages of these photodetectors
 include the voltage control of their spectral characteristics (at low temperatures), a relatively high
responsivity and detectivity (especially in the photodetectors with a relatively large number of GLs),
a high speed of operation,
and the possibility of operation in the frequency range (from 6 to 10 THz), where using more conventional  materials (e.g., A$_3$B$_5$ compounds) is hindered by the optical phonon absorption.

\bigskip

\begin{figure}[htbp]%%%Fig.1
\centering
\includegraphics[width=5.5cm]{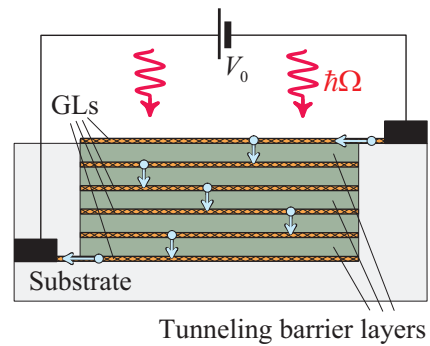}
\caption{Schematic structure of a vertical  I-GLPD with several (four) GLs.}
\end{figure}

\begin{figure}[htbp]%%%%Fig.2
\centering
\includegraphics[width=7.0cm]{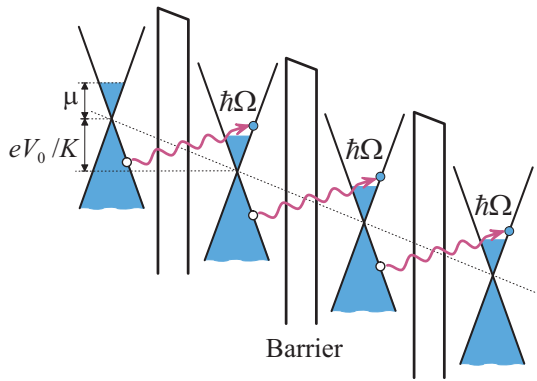}
\caption{Band diagram of 
a I-GLPD  under applied bias %voltage.
The arrows indicate the radiative interband transitions from  upper to lower GLs (which provide 
the main contribution to the photocurrent).}
\end{figure}

\section{Structure of I-GLPDs  and their operation principle}

Figure~1 shows the structure of the vertical intersubband GL-photodetector (I-GLPD) under consideration.
It  consists of  several 
n-doped GLs  separated by 
thin (tunneling-transparent) barrier layers of WS$_2$
or similar material. The bias voltage $V_0$ applied between the extreme GLs serving as the emitter (top GL) 
and collector (bottom GL), respectively.
Figure~2 shows the device band diagram under  bias  (several GL-structure periods).
As seen from Fig.2, the energy gaps between the GLs  are equal to 
$eV = eV_0/K$,
 where $K = 1,2,3,...$ is the number of the inter-GL barriers in the device.

Absorption by the  I-GLPD of the normally  incident photons with the energy $\hbar\Omega$ polarized 
in the GL plane, the absorption of a portion the photons is accompanied by
the electron transitions between the neighboring GLs (see the wavy arrows in Fig.~2). Such direct but inter-GL transitions produce
the inter-GL photocurrent.
At low temperatures, the spectral range, for the effective  interband inter-GL radiative transitions   
is given by the following inequalities: 

\begin{equation}\label{eq1}
2\mu - eV < \hbar\Omega < 2\mu + eV,
\end{equation}
where $\mu$ is the electron Fermi energy in the GLs determined by the donor density $\Sigma_i$ 
and (to some extent) by the device temperature $T$. For  the degenerate electron gas in the GLs ($k_BT \ll \mu$
where $k_B$ is the Boltzmann constant) $\mu \simeq \mu_i$, where

\begin{equation}\label{eq2}
\mu_i \simeq \hbar\,v_W\sqrt{\pi\Sigma_i}.
\end{equation}
Here $v_W \simeq 10^8$~cm/s is the characteristic velocity of electrons in GLs.

 The reverse transitions with the emission of a photon with the energy $\hbar\Omega$  are 
 suppressed (at sufficiently low temperatures) due to the Pauli blocking.  
 The interband intra-GL transitions are also possible. However,
at a certain relation between the photon energy $\hbar\Omega$, Fermi energy $\mu$, and
the inter-GL potential drop $V$, these transitions 
%(shown in Fig.2 by the crossed wavy arrows) 
can also be essentially blocked due to the Pauli
principle, particularly at low temperatures. This blocking mechanism is also effective for   the radiative transitions to the GLs 
with a higher potential energy, 
%(from the right to the left in Fig.2). 

 The interband transitions within the same GL  can result in  heating of the electron system~\cite{26}, 
 particularly at $\hbar\Omega > 2\mu$. However, due to a weak temperature dependence of the inter-GL 
 tunneling~\cite{26,27}, the pertinent contribution to the
inter-GL current  turns out to be relatively small if the Fermi energy $\mu$ is much smaller that
the barrier height (equal to the conduction band-offset, $\Delta_C$, between the GL and the barrier material), 
so that the thermionic emission over the barrier is insignificant. We exclude this case from 
our consideration here.

Due to the electric field in the top and bottom  GLs, the electron densities and, hence, the Fermi energies, 
$\mu_e$ and $\mu_c$, in these GLs differ from $\mu$. Taking into account the neutrality of the internal GLs,  at 
low relatively temperatures and at not too strong bias voltages,

\begin{equation}\label{eq3}
\mu_e \simeq \mu_i\biggl[1 + \frac{eV_0}{2(eV_i + \mu_i)}\biggr],
 \qquad 
\mu_c \simeq \mu_i\biggl[1 - \frac{eV_0}{2(eV_i + \mu_i)}\biggr].
\end{equation}
Here $V_i = (4\pi\,e\Sigma_idK/\kappa$, where $e$ is the electron charge, $d$ is the thickness
of the barrier layers, and $\kappa$ is  their dielectric constant.
However, a deviation of $\mu_e$ and $\mu_c$ from $\mu$ virtually does not affect the results obtained 
in the following and is disregarded.

\section{Photocurrent and responsivity}

Considering the absorption of normally incident radiation causing the vertical (conserving the electron momentum) interband inter-GL transitions,
the photocurrent density $j_{\Omega}$ , which is proportional to the difference in the rates of the electron 
transitions
from the upper  to the lower GLs and the reverse transitions, can be presented as

\begin{equation}\label{eq4}
j_{\Omega} = e\beta\theta\,G_{\Omega} I_{\Omega}
\end{equation}
Here $\beta = \pi\,e^2/\hbar\,c \simeq 0.023$ is the characteristic  (independent of the photon energy) coefficient of the interband absorption of normally incident electromagnetic radiation associated with the vertical transitions in GLs (see, for example,~\cite{15}), $c$ is the speed of light, 
$\theta \leq 1$ is the inter-GL overlap integral, $I_{\Omega}$ is the photon flux of the incident radiation 
($I_{\Omega}\hbar\Omega$ is the THz or IR power density),
and

$$
G_{\Omega} \simeq \frac{\displaystyle\rho\biggl(\frac{\hbar\Omega + eV}{\hbar\Omega}\biggr)\cdot\sinh\biggl(\frac{\hbar\Omega + eV}{2k_BT}\biggr)}
{\biggl[
\displaystyle\cosh\biggl(\frac{\hbar\Omega + eV}{2k_BT}\biggr) + 
\displaystyle\cosh\biggl(\frac{\mu}{k_BT}\biggr)\biggr]} 
$$

\begin{equation}\label{eq5}
- \frac{\displaystyle\rho\biggl(\frac{\hbar\Omega - eV}{\hbar\Omega}\biggr)\cdot\sinh\biggl(\frac{\hbar\Omega - eV}{2k_BT}\biggr)}
{\biggl[
\displaystyle\cosh\biggl(\frac{\hbar\Omega - eV}{2k_BT}\biggr) + 
\displaystyle\cosh\biggl(\frac{\mu}{k_BT}\biggr)
\biggr]} 
\end{equation}
is the voltage-dependent Pauli blocking factor, where $\rho(x) = x\Theta(x)$ and $\Theta(x)$ is the unity step function: $\Theta (x) = 0$ for $x < 0$ and  $\Theta (x) = 1$ for $x > 0$. The factors
$\rho\displaystyle\biggl(\frac{\hbar\Omega + eV}{\hbar\Omega}\biggr)$  and $\rho\displaystyle\biggl(\frac{\hbar\Omega - eV}{\hbar\Omega}\biggr)$ in the present form appear due to  the  linear dependences of the densities  of states on the energy in the valence and conductance bands.
Similar factor results in the photon energy dependence of the interband inter-GL absorption coefficient in GL structures at the transverse voltage in contrast to the interband intra-GL absorption coefficient 
(the latter is independent of the photon energy~\cite{15}).  In Eq.~(4) we have disregarded the reflection of the incident radiation from
the top of the GL-structure. The effect of the radiation reflection can easily accounted for by the proper renormalization of the quantity $I_{\Omega}$. 
We have also disregarded any "superlattice" effects, in particular, the formation of the energy gap because due to
the presence of the electric field in the barrier layers leading to the Stark-ladder  electron propagation. 
The applicability of Eq.~(4) for the I-GLPDs with  relatively large number of the barrier $K$ can be limited by the case when  the intensity of the radiation weakly decreases with its penetration  
into the depth of the GL-structure, i.e., by a relatively low absorption in each GL (see below).

 The quantity $\theta$ in Eq.~(4)  is a function of the spacing between GLs $d$ , see, Appendix A):
$\theta = e^{-2kd}(1 + kd)^2$, where $k = \sqrt{2m\Delta_C}/\hbar$,  $\Delta_C$ is the band offset,
and $m$ is the effective mass in the barrier. Assuming $\Delta_C = 0.4$~eV, $m = 0.27$ of 
the free electron mass~\cite{29}, and $d = 1.5$~nm, we obtain  $\theta \simeq 0.28$.
At low temperatures, the factor $G_{\Omega} \sim 1$ and $G_{\Omega} \ll 1$ inside and outside the interval 
given by inequalities (1), respectively.

Using Eqs.~(4) and (5), we arrive at the following expression for the GLPD responsivity 
${\cal R}_{\Omega} = j_{\Omega}/\hbar\Omega\,I_{\Omega}$ 
(current responsivity measured in the A/W units): 

$$
{\cal R}_{\Omega}
= 
{\overline {\cal R}_{\Omega}}\biggl\{\frac{\displaystyle\rho\biggl(\frac{\hbar\Omega + eV}{\hbar\Omega}\biggr)\cdot\sinh\biggl(\frac{\hbar\Omega + eV}{2k_BT}\biggr)}
{\biggl[
\displaystyle\cosh\biggl(\frac{\hbar\Omega + eV}{2k_BT}\biggr) + 
\displaystyle\cosh\biggl(\frac{\mu}{k_BT}\biggr)\biggr]} 
$$
\begin{equation}\label{eq6}
- \frac{\displaystyle\rho\biggl(\frac{\hbar\Omega - eV}{\hbar\Omega}\biggr)\cdot\sinh\biggl(\frac{\hbar\Omega - eV}{2k_BT}\biggr)}
{\biggl[
\displaystyle\cosh\biggl(\frac{\hbar\Omega - eV}{2k_BT}\biggr) + 
\displaystyle\cosh\biggl(\frac{\mu}{k_BT}\biggr)
\biggr]} \biggr\}
\end{equation}
with
\begin{equation}\label{eq7}
{\overline {\cal R}_{\Omega}} = \displaystyle\biggl(\frac{e\beta\theta}{\hbar\Omega}\biggr).
\end{equation}
 
 As seen from Eqs.~(6) and (7), the I-GLPD responsivity 
is independent of the number of GLs. Similar situation occurs in quantum-well infrared photodetectors (QWIPs) 
with the  photoexited and injected electrons moving perpendicular to
the multiple-QW structure~[30-32] (although the dependence on the number of QWs  occurs due to the damping of the radiation  caused by its absorption in the depth of the multiple-GL structure, the contact effects, 
and at elevated modulation frequencies  and radiation powers~\cite{31,33,34}).

\begin{figure}[t]
\centering
\includegraphics[width=6.5cm]{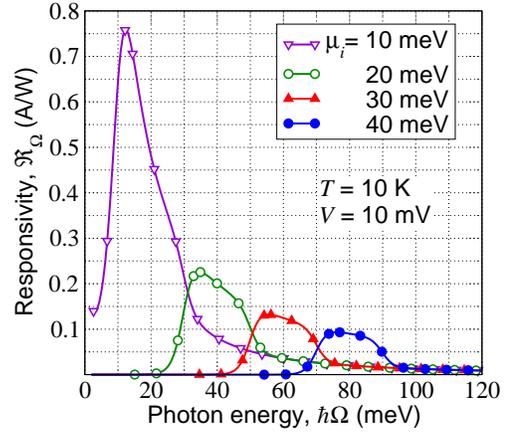}%F3
\caption{Spectral dependences of the responsivity ${\cal R}_{\Omega}$ of I-GLPDs with different
electron Fermi energies $\mu_i = 10 - 40$  at
$T = 10$K and $V = 10$~mV.}
\end{figure}

\begin{figure}[t]
\centering
\includegraphics[width=6.5cm]{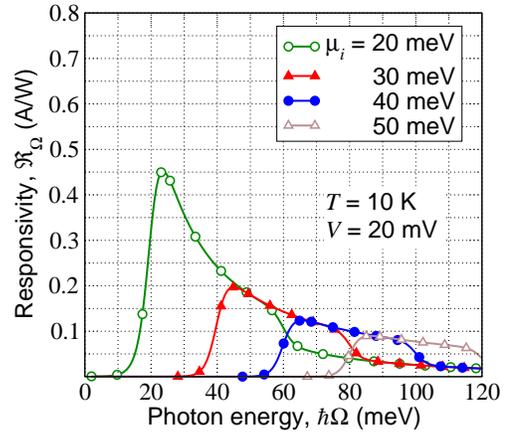}%F4
\caption{The same as in Fig.~3 but for
 $\mu_i = 20 - 50$ and  $V = 20$~mV.}
\end{figure}

At  low temperatures $k_BT \ll  \hbar\Omega, eV $, Eq.~(6) yields the spectral dependences
with a pronounced maximum in the range $2\mu_i - eV < \hbar\Omega < 2\mu_i + eV$.
In particular,
in the limit $T \rightarrow 0$,
one obtains
${\cal R}_{\Omega} = {\overline {\cal R}_{\Omega}}$  if $2\mu_i - eV < \hbar\Omega < 2\mu_i + eV$ and 
${\cal R}_{\Omega} = 0$  if $\hbar\Omega < 2\mu_i - eV$ or  $\hbar\Omega >2\mu_i + eV)$.

Figures~3 and 4 show the spectral dependences ${\cal R}_{\Omega}$ calculated using Eq.~(6)
for the I-GLPDs
with a WS$_2$ barrier of the thickness $d = 1.5$~nm  ($\theta \simeq 0.28$),
for different Fermi energies: $\mu_i = 10 - 40$~meV 
at $T = 10$~K and $V = V_0/K = 10$~mV and 20~mV.  
As seen,  the spectral dependences of the responsivity at $T = 10$~K
exhibit relatively narrow virtually symmetrical (at $V = 10$~mV) and markedly asymmetrical
(at $V = 20$~mV) peaks. The position centers of these peaks $\hbar\Omega_c$ 
are determined by  the Fermi energy ($\hbar\Omega_c \simeq 2\mu_i$), while
 their widths
$\hbar\Delta\Omega$ is determined by the bias voltage ($\hbar\Delta\Omega \simeq 2eV$).
The positions of the peak maxima $\hbar\Omega_m$ are shifted toward smaller photon energies
($\hbar\Omega_m \leq \hbar\Omega_c$).
As it was pointed out above, a steep roll-off of the responsivity at $\hbar\Omega < eV - \mu_i$
and $\hbar\Omega > eV - \mu_i$ is due to the Pauli blocking of the interband inter-GL radiative transitions 
outside the indicated photon energy range. Such a blocking is essential in
the cases corresponding to Fig.~3 because of a pronounced degeneracy of the electron gas in GLs
and a steep variation of the electron distribution function
at chosen values of the Fermi energy and the temperature.

\begin{figure}[t]
\centering
\includegraphics[width=6.5cm]{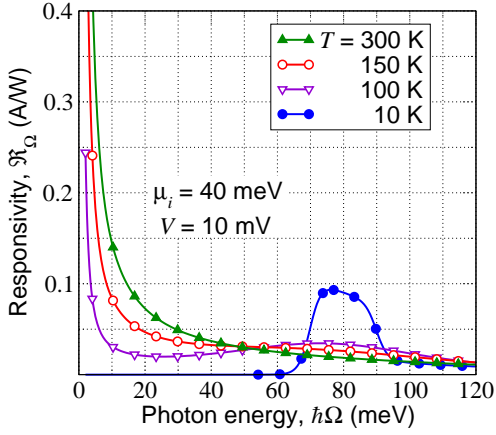}
\caption{Spectral dependences of the GLPD responsivity ${\cal R}_{\Omega}$ 
at different temperatures   for $\mu_i = 40$~meV and $V = 10$~mV.
}
\end{figure}

A  temperature increase leads to a substantial smearing of the electron distribution function and
the responsivity peaks.
As a result,  the spectral characteristics of the I-GLPD responsivity become monotonic functions of the photon energy at sufficiently high temperatures.
Indeed, 
at relatively high temperatures  ($\hbar\Omega , eV < k_BT $), from Eq.~(6) we obtain

\begin{equation}\label{eq8}
{\cal R}_{\Omega} \simeq \frac{\overline{\cal R}_{\Omega}}{[1 + \cosh (\mu_i/k_BT)]}
\biggl(\frac{eV}{k_BT}\biggr) 
\end{equation}
with

\begin{equation}\label{eq9}
{\cal R}_{\Omega} \simeq \overline{\cal R}_{\Omega}
\biggl(\frac{2eV}{k_BT}\biggr) 
\end{equation}
at $\hbar\Omega , eV, \mu_i < k_BT $, and

\begin{equation}\label{eq10}
{\cal R}_{\Omega} \simeq \overline{\cal R}_{\Omega}\exp\biggl(- \frac{\mu_i}{k_BT}\biggr)
\biggl(\frac{2eV}{k_BT}\biggr) 
\end{equation}
at  $\hbar\Omega , eV < k_BT < \mu_i$.
 As follows from Eqs.~(8) - (10), at  elevated  temperatures,
 ${\cal R}_{\Omega}  \propto 1/\hbar\Omega$ (via the dependence of $\overline{\cal R}_{\Omega}$ on the photon energy). 
 
Figure~5 shows the transformation  of the responsivity vs. photon energy dependences with increasing
temperature: the smearing of the responsivity maxima (compare the curves for
$T = 10$~K and 100~K) and transition to the monotonic dependences given by
Eq.~(10). As seen  from Fig.~5, at the low end of the photon spectrum, the responsivity increases with the temperature
in agreement with Eq.~(10). 

According to Figs.~3 - 5, I-GLPDs exhibit fairly different spectral characteristics  
at low and elevated temperatures with a rather high values of the responsivity in both temperature ranges.

\section{Dark current and dark current limited detectivity}

\begin{figure}[t]
\centering
\includegraphics[width=6.5cm]{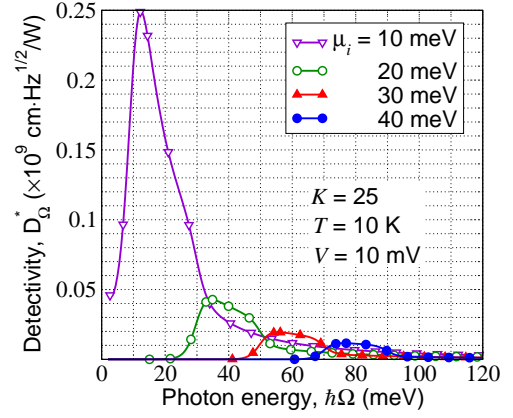}%%%%Fig.6
\caption{Spectral dependences of the GLPD detectivity $D^*_{\Omega}$ 
for different $\mu_i$,
$T = 10$~K, and $V = 10$~mV
(i.e., for the same parameters as in Fig.~3), and  $K = 25$.
}
\end{figure}

\begin{figure}[t]
\centering
\includegraphics[width=6.5cm]{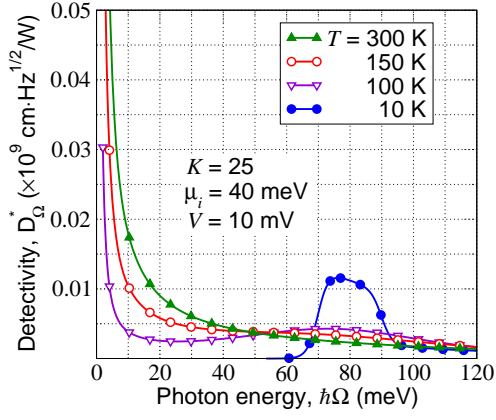}%%%%Fig.7
\caption{Spectral dependences of the GLPD detectivity $D^*_{\Omega}$ 
at different temperatures, $K = 25$, and the same other parameters as in Fig.~5.
}
\end{figure}

Taking into account the Fermi-Dirac statistics of electrons and their linear
dispersion relation in GLs, the electron tunneling
current between the neighboring GLs at $k_BT < eV, \mu_i$ (a strong degeneracy of the electrons in GLs) can be presented in the following simplified form:

\begin{equation}\label{eq11}
j_{dark} = \frac{e[\mu_i^2 - (\mu_i - eV)^2]}{\pi\hbar^2v_W^2\tau_{esc}} 
\simeq \frac{2e\mu_ieV}{\pi\hbar^2v_W^2\tau_{esc}}. 
\end{equation}
In the opposite case  ($k_BT > eV, \mu_i$), one obtains
\begin{equation}\label{eq12}
j_{dark} \simeq \frac{\pi\,e^2k_BTV}{6\hbar^2v_W^2\tau_{esc}},
\end{equation}
The quantity   $\tau_{esc}$ in Eqs.~(11) and (12)
is the characteristic escape time from one GL to another.
Due to the non-alignment of the Dirac points at the applied voltage, such transitions
are possible if they are accompanied by the variation of the electron momentum due to
the electron scattering caused by  disorder~\cite{26,27}. 
We roughly estimate  the escape time as $\tau_{esc} = \tau/\theta$,
where $\tau$ is the electron momentum relaxation time.
As follows from Eqs.~(11) and (12), the dark current is a slow (non-exponential) function of the temperature.
Some reinforcement of the temperature dependence in comparison with that given by Eq.~(10) might 
arise from a decrease in $\tau_{esc}$ with increasing $T$ (due to a decreasing 
 $\tau$ vs $T$ dependence).
As an example, setting  $\mu_i = eV = 10$~meV and 
$\tau = 10^{-12} $~s and  assuming for WS$_2$ barriers $\theta = 0.28$,
$ \tau_{esc} = 3.6\times 10^{-12} $~s (see Appendix B), 
at low temperatures and at $T = 300$~K
we obtain from Eqs.~(11) and (12)
$j_{dark} \sim 7\times 10^2 $~A/cm$^2$ and 
$j_{dark} \sim 14\times 10^2 $~A/cm$^2$, respectively.

Considering that the noise current is given by $J_{noise} = \sqrt{4egJ_{dark}\Delta f}$,
where $\Delta f$ is the bandwidth,
 the photodetector dark current limited  detectivity can be calculated using the following formula: 

\begin{equation}\label{eq13}
D_{\Omega}^* = \frac{{\cal R}_{\Omega}}{J_{noise}}\sqrt{A\cdot\Delta f} 
=\frac{{\cal R}_{\Omega}^J}{\sqrt{4egj_{dark}}}, 
\end{equation}
where $A$ is the device area and $g$ is the current gain (both photoelectric and dark current gain). In the I-GLPDs under
consideration, in which the transitions occur only between the neighboring GLs, $g = K^{-1}$ 
(see, for example, ~\cite{34}).  Taking into account that ${\cal R}_{\Omega} \propto \beta\theta $,
 $j_{dark} \propto \theta$, and $g  = K^{-1}$, we obtain (for fixed $V = V_0/K$)

\begin{equation}\label{eq14}
D_{\Omega}^* \propto \beta\sqrt{\theta\,K}.
\end{equation}

Figures~6 and 7 show the detectivity of the I-GLPDs with different values of the Fermi energy $\mu_i$
(different donor densities
$\Sigma_i$) at different temperatures $T$ calculated 
using Eq.~(6) with Eqs. (11) - (13). One can see that the spectral dependences of $D^*_{\Omega}$ qualitatively repeat those of ${\cal R}_{\Omega}$.

\begin{figure}[t]
\centering
\includegraphics[width=6.5cm]{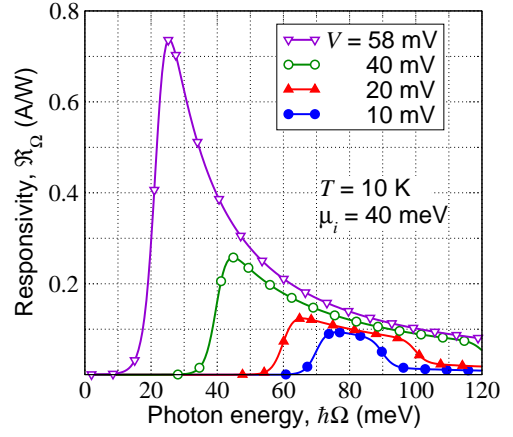}%%%%Fig.8??????????????????
\caption{The GLPD responsivity ${\cal R}_{\Omega}$ vs, photon energy $\hbar\Omega$ at different bias voltages $V$
($T = 10$~K and $\mu_i = 40$~meV).
}
\end{figure}

\begin{figure}[t]
\centering
\includegraphics[width=6.5cm]{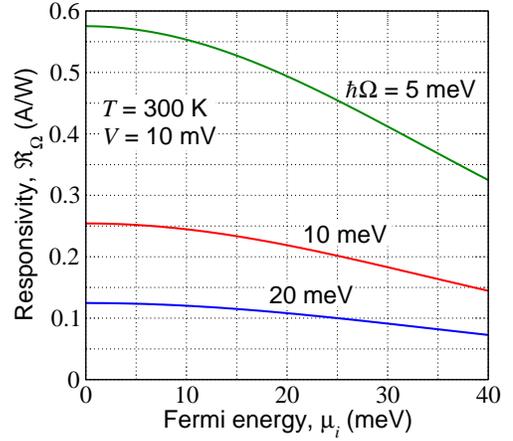}%%%%Fig.9
\caption{Dependences of the GLPD responsivity ${\cal R}_{\Omega}$
  on the Fermi energy $\mu_i$ 
calculated for different photon energies $\hbar\Omega$ ($T = 300$~K and $V = 10$~mV).
}
\end{figure}
\begin{figure}[th]
\centering
\includegraphics[width=6.5cm]{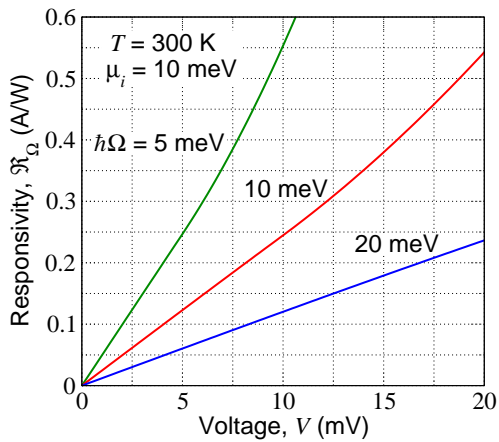}%%%%Fig.10
\caption{ of the GLPD responsivity ${\cal R}_{\Omega}$
 on bias voltage $V$ 
calculated for different photon energies $\hbar\Omega$ ($T = 300$~K and $V = 10$~mV).
}
\end{figure}

\section{Manipulation of the I-GLPD characteristics and their optimization}

\begin{figure}[t]
\centering
\includegraphics[width=6.5cm]{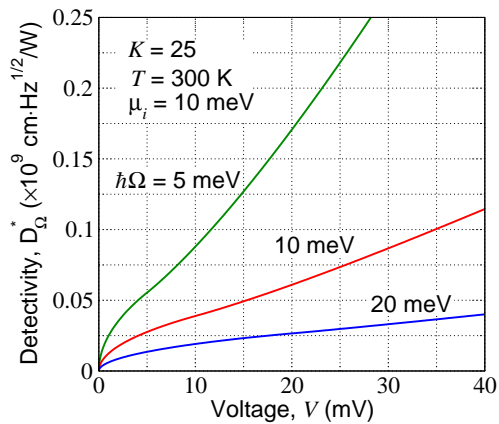}%%%%Fig.11
\caption{Dependences of the GLPD detectivity
 $D^*_{\Omega}$ on bias voltage $V$
calculated for different photon energies $\hbar\Omega$ ($K = 25$, $T = 300$~K, and $V = 10$~mV).
}
\end{figure}

\begin{figure}[t]
\centering
\includegraphics[width=6.5cm]{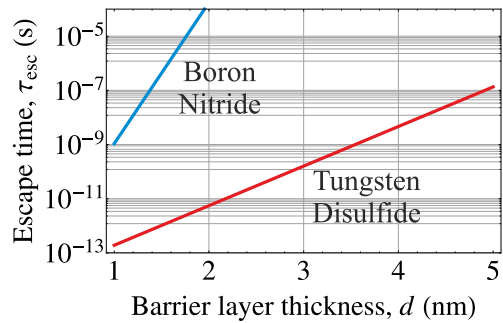}
\caption{Inter-GL escape time vs. barrier thickness for hBN barrier (extracted from the experiment~\cite{5,41}) and for WS$_2$ barrier (recalculated using the data for hBN barrier). 
}
\end{figure}
As follows from the above formulas for the I-GLPD responsivity and detectivity, these quantities are determined by the Fermi energy $\mu_i$ (doping level of GLs), the energy of the incident photons $\hbar\Omega$,
the temperature $T$, and the bias voltage $V_0$. 
This provides an opportunity to optimize the I-GLPD characteristics for different application.
According to Figs.~3 -7, the spectral
dependences of the responsivity and detectivity at low temperatures (at about $T = 10$~K and lower) are
located in rather narrow ranges of the photon energy, which are determined by the Fermi energy, i.e.,
by the donor density in GLs. It opens up the prospect to create GL-based multicolor photodetectors,
particularly those operating the spectral range where the operation of the
detectors based on A$_3$B$_5$ materials is hindered by the optical phonon absorption.
The width of spectral peaks can be decreased by using lower bias voltages (down to $V \sim k_BT/e$,
 i.e., down to $V \sim 1$~mV at $T = 10$~K).
An increase in the bias voltage leads to lower dark current [see Eq.~(11)] and, hence, in a rise in the detectivity. Figure~8 shows a pronounced rise in the peak  responsivity with increasing bias voltage.
As seen, an increase in the voltage leads also to a marked transformation
of the spectral characteristic and a pronounced  shift of the peak position toward smaller photon energies.
 
A pronounced variation of the I-GLPD responsivity and detectivity at elevated temperatures
also opens up the possibility of the characteristics manipulation and the photodetector optimization. Figures~9 and 10 show the room temperatures responsivity versus the Fermi energy and the bias voltage.
Figure~11 shows the I-GLPD detectivity as a function of the bias voltage. Comparing Fig.~9 and 10,
one can see that while the responsity is a linear function of the bias voltage (in the voltage range
under consideration), the detectivity exhibits a sub-linear behavior. This is because of
an increase in the dark current with increasing voltage.

Although the I-GLPD responsivity is practically independent of the number of the inter-GL barriers $K$
(and the number of GLs $(K - 1)$) in the device structure if this number is not too large,
an increase in this number leads to an marked  enhancement of the I-GLPD detectivity.
The dependence of ${\cal R}_{\Omega}$ on $K$ appears  when the intensity of the incident radiation
decreases as the radiation  is absorbed by the GLs which are closer to the illuminated device surface.
At low temperatures when the Pauli blocking is pronounced and $2\mu_i - eV < \hbar\Omega < 2\mu_i$,
such effect becomes pronounces when $\beta\theta\,K$ approaches unity.
It implies, that the reasonable
value of $K$ is $K \leq 1/\beta\theta \simeq 150$. When $K > 1/(\beta\theta)$, $D^*_{\Omega}$ as a function of $K$ saturates. At elevated temperatures, the interband intra-GL absorption is essential, so that the pertinent limitation reads $K \simeq 1/\beta \simeq 40$. At marked absorption of the radiation when $K$ is large,
the electric fields in the barrier layers can be somewhat different (larger in the barrier layers remote from
the irradiated surface). This can result in some modification of the obtained formulas (however, not leading to  significant changes of the obtained characteristics)  which is out of the scope of this work.

\section{Comparison with quantum-well  and some other photodetectors}

The I-GLPD structure resembles the structures of quantum-well infrared photodetectors (QWIPs)
and THz quantum-well detectors (THz-QWDs) with multiple QWs. Although QWIPs and THz-QWDs use mainly
the photoexcitation from the bound states to the continuum states (in contrast to I-GLDs exploiting
the inter-GL tunneling), it makes sense to compare their characteristics.
As follows from the obtained results, 
I-GLPDs at low temperatures ($T = 10$~K)and photon energy $\hbar\Omega \sim 80$~meV exhibit the responsivity ${\cal R}_{\Omega}$  on the order of 0.1~A/W (see Figs.~3 and 4).

At the photon energy $\hbar\Omega \simeq 13.5$~meV (corresponding to the radiation wavelength $\lambda \simeq 87~\mu$m)
in the temperature range $150 - 300$~K,   GaAs-AlGaAs THz-QWLs studied in Ref.~\cite{35}
exhibit ${\cal R}_{\Omega} \simeq 0.009 - 0.018$~A/W at $T = 10$~K, whereas  I-GLPDs  with the parameters corresponding to Figs.~3 and 5 exhibits
${\cal R}_{\Omega} \simeq 0.75$~A/W and 0.25~A/W at $T = 10$~K and 300K, respectively. 
The detectivity of the latter THz-QWDs
based on the 60-QW structure  at $\hbar\Omega \simeq 13.5$~meV  and $T = 10$~K was reported to be $D^*_{\Omega} \sim 5\times 10^7$~cm~$\cdot$~Hz$^{1/2}$/W. As  seen from Fig.~6, the I-GLPD detectivity   
at the same photon energy and the temperature reaches the value of $D^*_{\Omega} \sim 25\times 10^7$~cm~$\cdot$~Hz$^{1/2}$/W, i.e, five times higher.

However, an increase in the bias voltage to $V = 40$~mV (see Fig.~10)  and an increase in $K$ from $K = 20$ to 60 yields the same $D^*_{\Omega}$ (at $T = 300$~K) as in the
THz-QWD in question (at $T = 10$~K). The responsivities of  a GaAs-AlGaAs THz-QWD~\cite{36} 
and an I-GLPD (see Fig.~8)
for $\hbar\Omega \simeq 28$~meV ($\lambda = 42 ~\mu$m or $\Omega/2\pi = 7.1$~THz) and $T = 10$~K can be  close to each other depending on the voltage and can be  equal to  ${\cal R}_{\Omega} \simeq  0.5 - 0.6$~A/W.

At room temperature in the range $\hbar\Omega = 20 - 40$~meV,  covering the photon energy range which is  not accessible by A$_3$B$_5$
based detectors due to optical phononabsorption (the region from 33 to 37~meV for GaAs-AlGaAs devices~\cite{36}), I-GLDs could demonstrate rather reasonable values of the  responsivity about of 30 - 75~mA/W with a modest detectivity on the order of $0.5 - 0.9)\times10^7$~cm~$\cdot$~Hz$^{1/2}$/W(see Figs.~5
and 7).
 
Due to the possibility of  a relatively high-speed operation of  QWIPs, they   are considered as candidates for the communication systems with the modulation frequencies in the sub-THz range.
The QWIP responsivity as a function of the modulation frequency is mainly determined
by the electron transit time across the structure and the probability of the electron capture
into the QWs. The theoretical estimates and experimental data show the possibility of the QWIPs effective operation up to about hundred GHz~\cite{34,37,38}. However in such a range of the modulation frequencies   the  photoelectric gain is suppressed, so that the QWIP responsivity is much smaller than at  low modulation frequencies.

One of the most remarkable features of the I-GLDs is their high-speed operation associated with 
short inter-GL tunneling times. 
There are different complex approaches to determine the characteristic tunneling time (see, for example, ~{37,38} and the speed of
devices using the tunneling effects. We assume that the speed of the I-GLPD operation, i.e., the maximum frequency of the radiation modulation $\omega$ should be at least much smaller than
$\omega_m = 1/\tau_t$, where $\tau_t$ is the inter-GL tunneling time.   Following to Buttiker and Landauer~\cite{39,40},
we assume that $\tau_t = d\sqrt{m*/2\Delta_c}$, where $m*$ is the electron effective mass in the barrier. For the WS$_2$ barrier width $d$ of 1.32~nm,  setting $\Delta_C = 0.4$eV and $m* = 0.27$
of the free electron mass, we obtain $\tau_t \simeq 2.3$~fs. This implies that 
$\omega_m/2\pi \simeq 70$~THz, so  that the I-GLPDs can operate up to several THz  modulation frequencies. 
For high speed applications, it is  desirable to provide a  I-GLPD impedance of $Z =50$~Ohm.
Using Eqs.~(11) and (12) and the consequent estimates for low and room temperatures, respectively, we obtain $Z \simeq (7 - 14)\times10^{-6}K/A$~Ohm, where $A$ is the I-GLPD area. Therefore, to match the 50~Ohm impedance, 
for the number of the inter-GL barriers $K = 25$, one needs $A = (3.5 - 7)\times10^2 \mu$m$^2$.

The THz- and IR-PDs based on the  GL structures with lateral p-i-n junctions (called the lateral p-i-n GLPDs), exploiting the inter-band intra-GL photoexitation of electrons and holes, considered previously~\cite{18,23} can exhibit a substantially higher responsivity. However,
the speed of their operation is limited (below 25 - 50~GHz) by relatively long electron and hole transit time
between the p- and n-regions. Hence,  the vertical I-GLPDs  can surpass the lateral p-i-n GLPDs in speed
in the responsivity can surpass the latter in the speed of operation, even though they have a smaller responsivity.

Thus, the I-GLPDs exhibit sufficiently high responsivity and detectivity at room temperature and can surpass other THz- and IR-PDs in the operation speed. One can anticipate the applications of the I-GLDs in analog transmission systems
with the THz or mid-IR  carrier frequencies and sub-THz or even THz modulation frequencies. 

\section{Conclusions}
We propose   the cascade vertical I-GLPDs    based on  
  multiple-GL structures with thin  tunnel-transparent barrier layers and exploiting the interband inter-GL radiative transitions. These devices should be  able to operate in the THz and mid-infrared spectral ranges. 
  Using the developed device model, we calculated the I-GLPD responsivity and detectivity
  as functions of the photon energy, the bias voltage, and  the number of GLs in the structure
in a wide range of temperatures (from cryogenic to room temperatures) and evaluated the I-GLPD speed of operation.
We demonstrated that the I-GLPD characteristics strongly depend on the GL doping level and can be effectively controlled by the bias voltage.
The I-GLPD can exhibit a sufficiently high responsivity (about several tenth of A/W) 
both at low and room temperatures  and a reasonable 
detectivity at room temperature, surpassing or competing with
other THz and IR-photodetectors. Due to the tunneling origin of the photocurrent and dark current,
the THz I-GLPDS and mid-IR I-GLPDs  could achieve a higher speed of operation than
the existing photodetectors. These new devices can  be used in the analog optical communication systems with the sub-THz and THz modulation
frequencies.

\section*{Appendix A. Inter-GL overlap integral}

The envelope wave functions, $\psi_l(z)$ and $\psi_r(z)$,  depending on the coordinate $z$ in the direction perpendicular to the GL plane for two neighboring GLs can be presented as

\begin{equation}\label{eq15}
\varphi_l(z) = \frac{\Phi_l}{\sqrt{\int^{\infty}_{-\infty}|\Phi_l|^2dz}}, \qquad
\varphi_r(z) = \frac{\Phi_r}{\sqrt{\int^{\infty}_{-\infty}|\Phi_r|^2dz}}.
\end{equation}
Here
\begin{equation}\label{eq16}
\Phi_l(z) = \exp(-k|z + d/2|), \qquad \Phi_r(z) = \exp(-k|z - d/2|),  
\end{equation}
where
$k = \sqrt{2m\Delta_C}/\hbar$, $\Delta_C$ is the conduction band offset between the barrier
material  and GLs,
and $m$ is the effective mass in the barrier.
Neglecting the overlap of the wave functions of the distant GLs,
the probability of the photon absorption accompanied by the electron transition
between the valence band in one GL to the conduction band in the neighboring one can be presented as
\begin{equation}\label{eq17}
\beta_{inter-GL} = \frac{\pi\,e^2}{\hbar\,c} \theta = \beta\theta,  
\end{equation}
where
\begin{equation}\label{eq18}
\theta= \biggl|\int_{-\infty}^{\infty}\varphi_l(z)\varphi_r(z)\biggr|^2 = e^{-2kd}(1 + kd)^2  
\end{equation}
is the overlap integral and $\beta = \pi\,e^2/\hbar\,c \simeq 0.23$.

\section*{Appendix B. Escape time}

Using the experimental data from~\cite{5}, one can  estimate the escape time  in graphene-BN-graphene structure. Knowing  the peak (resonant) current $I_p=35$ nA at the bias voltage $V=0.3$ V,  the device area $A=0.3~\mu$m$^2$, and   the electron and hole  densities   $\Sigma_e=\Sigma_h =1.8\times {10^{12}}$~cm$^{-2}$, we obtain the inter-GL tunneling escape time  $ \tau^{\prime}_{esc} = 5\times 10^{-8}$~s.
To estimate the inter-GL tunneling escape time, $\tau_{esc}$, in graphene-WS$_2$-graphene structure, we 
take into account that according to  the Bardeen tunneling Hamiltonian approach, $\tau_{esc} \propto {e^{2\ae\, d}}$,where $\ae =\sqrt{2 m^{*} \Delta }/\hbar $,$d$ and $\Delta_C $ are the thickness and the height of the barrier, and $m^{*}$ is the effective mass in the intermediate (barrier) material.
Thus, the ratio of tunneling escape times  $\tau_{esc}^{\prime}$ and $\tau_{esc}$ in BN- and WS$_2$- structures can be presented as
\begin{equation}
\frac{\tau_{esc}^{\prime}}{\tau_{esc}}=\exp[2(\ae^{\prime} d^{\prime} - \ae d)].
\end{equation}

From the measurements of thickness-dependent resistivity of BN tunnel barriers we know $\ae^{\prime}=6$ nm$^{-1}$~\cite{41}. The obtained tunneling escape time  $\tau^{\prime}_{esc} = 5\times 10^{-8}$~s corresponds to four layers of hBN, which converts to $d^{\prime} = 4\times 0.33$ nm $ = 1.32$ nm. Known also graphene-WS$_2$ band structure parameters $\Delta_C = 0.4$ eV, $m_2^{*}=0.28{m_e}$, we can plot the dependence  of  $\tau_{esc}$ on thickness $d$ shown in Fig.~12. As seen from Fig.~12, at $d = 1.32$~nm
we obtain $\tau_{esc} \simeq 2\times 10^{-12}$~s, i.e., the value close to that used in the main text.
Since our way of extracting $\tau_{esc}$ provides only the leading exponent, hence the agreement between two approaches is quite reasonable.

\section*{Acknowledgments}

This work was supported by the Japan Society for promotion of Science (Grant-in-Aid for Specially Promoting Research $\# 23000008$),  Japan.  V. R., M. R., and D.S. acknowledge 
the support by the Russian Scientific Foundation (Project $\#14 − 29 − 00277$). The work by A.D.  
was also supported by 
the Dynasty Foundation, Russia. The work at the University
at Buffalo was supported by the National  Science Foundation TERANO and the Air Force Office of Scientific Research
grants. The work at RPI was supported by the US Army Research Laboratory 
Cooperative Research Agreement.

\end{document}